\documentclass[reqno,11pt]{article}
\usepackage{ifpdf}
\usepackage[usenames,dvipsnames]{xcolor}
\usepackage{ae}
\usepackage[T1]{fontenc}
\usepackage[ansinew]{inputenc}
\usepackage{mathrsfs}
\usepackage{amsmath}
\usepackage{amssymb}
\usepackage{graphicx}
\usepackage{color}
\definecolor{darkblue}{cmyk}{0.9,0.9,0,0}
\definecolor{darkgreen}{rgb}{0,0.55,0}

\usepackage{hyperref}

\usepackage{epsfig}
\usepackage{amsmath}

\newcommand{\comment}[1]{}

\newcommand{\beq}{\begin{equation}}
\newcommand{\eeq}{\end{equation}}
\newcommand{\beqq}{\begin{equation*}}
\newcommand{\eeqq}{\end{equation*}}
\newcommand\beqa{\begin{eqnarray}}
\newcommand\eeqa{\end{eqnarray}}
\newcommand\beqaa{\begin{eqnarray*}}
\newcommand\eeqaa{\end{eqnarray*}}
\newcommand\bea{\begin{array}}
\newcommand\eea{\end{array}}

\def\XXint#1#2#3{{\setbox0=\hbox{$#1{#2#3}{\int}$ }
\vcenter{\hbox{$#2#3$ }}\kern-.5\wd0}}

\def\XXint#1#2#3{{\setbox0=\hbox{$#1{#2#3}{\int}$}
\vcenter{\hbox{$#2#3$}}\kern-.5\wd0}}

\newcommand{\nn}{\nonumber}

\newcommand{\neqa}{\nonumber\end{eqnarray}}
\newcommand{\la}[1]{\label{#1}}

\newcommand{\eq}[1]{(\ref{#1})}

\newcommand{\Tr}{{\rm Tr}}

\newcommand{\hs}{\frac{\sqrt{3}}{2}}
\renewcommand{\d}{\partial}

\newcommand{\<}{{\langle}}
\renewcommand{\>}{{\rangle}}

\newcommand{\cL}{{\cal L}}

\newcommand{\re}{\relax{\rm I\kern-.18em R}}

\renewcommand{\sp}{p\hspace{-.40em}/}

\def\su2{{SU(2)}}

\def\[{\left[}
\def\]{\right]}

\def\s{\sigma}

\def\({\left(}
\def\){\right)}
\def\[{\left[}
\def\]{\right]}

\def\<{\langle}
\def\>{\rangle}

\def\cO{{\cal O}}

\def\mC{{\mathbb C}}

\def\s*{\ *_{\!\!\!\!\!\!\!\!\!\,_{\,_\text{\scriptsize{sym}}}}}
\def\hs*{\ \hat{*}_{\!\!\!\!\!\!\!\!\!\,_{\,_\text{\scriptsize{sym}}}}}
\def\d{\partial}

\def\i2{\frac{i}{2}}

\def\bQ{{\bf Q}}
\def\bP{{\bf P}}

\def\spi{\relax{\rm \pi\kern-0.5em /}}
\def\sA{\relax{\rm A\kern-0.5em /}}
\def\sp{\relax{\rm p\kern-0.5em /}}
\def\sd{\relax{\rm \d\kern-0.5em /}}
\def\sk{\relax{\rm k\kern-0.5em /}}
\def\sn{\relax{\rm n\kern-0.5em /}}
\def\sl{\relax{\rm l\kern-0.5em /}}
\def\sP{\relax{\rm P\kern-0.7em /}}
\def\sBethe{\relax{\rm \Bethe\kern-0.5em /}}
\def\cN{{\cal N}}

\newcommand{\es}{\emptyset}
\newcommand{\bes}{{\bar{\emptyset}}}
\newcommand{\vint}{\int_{{\bf |}}}

\newcommand{\ket}[1]{| #1 \rangle}

    \newcommand{\br}[1]{{\langle\!\!\langle} #1 {\rangle\!\!\rangle}}

\numberwithin{equation}{section}

\usepackage{varioref}
\usepackage{makeidx}
\makeindex

\usepackage[english]{babel}

\begin{document}

\begin{titlepage}

\setcounter{page}{0}
\renewcommand{\thefootnote}{\fnsymbol{footnote}}

%\begin{flushright}
%Durham, UK 
%\end{flushright}

\vspace{1cm}

\begin{center}

\textbf{\large\mathversion{bold} A review of the AdS/CFT Quantum Spectral Curve}

\vspace{1cm}

{\large Fedor Levkovich-Maslyuk \footnote{{\it Email:\/}
{\ttfamily fedor.levkovich at gmail.com}}\footnote{Also at Institute for Information Transmission Problems, Moscow 127994, Russia}
} 

\vspace{1cm}

\it Departement de Physique, Ecole Normale Superieure / PSL Research University, CNRS, \\ 24 rue
Lhomond, 75005 Paris, France

\vspace{1cm}

{\bf Abstract}
\end{center}
\vspace{-.3cm}
\begin{quote}
We give an introduction to the Quantum Spectral Curve in AdS/CFT. This is an integrability-based framework which provides the exact spectrum of planar $\cN=4$ super Yang-Mills theory (and of the dual string model) in terms of a solution of a Riemann-Hilbert problem for a finite set of functions. We review the underlying QQ relations starting from simple spin chain examples, and describe the special features arising for AdS/CFT. We also discuss the recently found links between the Quantum Spectral Curve and the computation of correlation functions. To appear in a special issue of J. Phys. A based on lectures given at the Young Researchers Integrability School and Workshop 2018.

\vfill
\noindent %July 2015

\end{quote}

\setcounter{footnote}{0}\renewcommand{\thefootnote}{\arabic{thefootnote}}

\end{titlepage}

\tableofcontents

\section{Introduction}

From its early days, the AdS/CFT correspondence has attracted great attention as it gives a hope to understand strongly coupled field theories by invoking tools from string theory. The best studied and most famous example is the holographic duality between $\cN=4$ super Yang-Mills theory (SYM) and string theory on ${\rm AdS}_5 \times {\rm S}^5$ \cite{Maldacena:1997re,Witten:1998qj,Gubser:1998bc}. Remarkably, it appears that this case may in fact be exactly solvable at least in the planar (large $N_c$) limit, which gives a hope to better understand both the mechanisms behind holography and the dynamics of 4d gauge theories in general.

One major outcome of the integrability program is a concise description of the exact planar spectrum of $\cN=4$ SYM in terms of a finite set of functional equations known as the Quantum Spectral Curve (QSC) \cite{Gromov:2013pga}, which will be the main focus of this review. Exploration of integrability started indeed with the spectral problem when it was observed in \cite{Minahan:2002ve} that the 1-loop spectrum is captured by an integrable spin chain. This triggered a great interest in the subject, and we refer the reader to \cite{Beisert:2010jr} where the development of various branches of the field is described in detail\footnote{We also refer to \cite{Bombardelli:2016rwb} for a recent review of the more standard integrability tools oriented towards AdS/CFT applications.}. The combination of approaches starting from classical integrability for the dual string \cite{Bena:2003wd}, the exact S-matrix and asymptotic Bethe ansatz \cite{Beisert:2005fw} led to a set of TBA/Y-system equations \cite{Gromov:2009tv,Bombardelli:2009ns,Arutyunov:2009ur,Gromov:2009bc} that capture, at least in principle, the full spectrum scaling dimensions for local single trace operators of the type $\cO(x)={\rm Tr}(\Phi_1(x)\Phi_2(x)\dots)$ where $\Phi_i$ are fundamental fields of the theory. While these equations have led to highly important results, they involve an infinite set of unknown functions, preventing their efficient solution and making precise calculations of the spectrum rather difficult. It took several more years to reformulate them in terms of a finite system \cite{Gromov:2011cx}, and finally in 2013 they were reduced to the highly compact and transparent form which goes by the name of Quantum Spectral Curve.

The QSC appears to be the ultimate solution of the spectral problem. It has the form of a set of functional equations for a only a few unknown functions, which have a rather clear algebraic meaning. The QSC is based on a universal algebraic framework known as QQ relations (or Q-system) \cite{Kulish:1979cr,Kulish:1985bj,Krichever:1996qd,Dorey:2006an,Kazakov:2007fy}, which can be formulated for a wide variety of integrable models including standard integrable spin chains. These essentially generalize the Baxter difference equation to the case of more involved symmetries, covering in particular the $psu(2,2|4)$ global symmetry of $\cN=4$ SYM. The unknowns in these equations are the Q-functions which for spin chains are simply polynomials. However, for $\cN=4$ SYM the Q-functions develop a nontrivial analytic structure, which plays a key role in the construction. These functions have a prescribed set of branch cuts and live on an infinite-sheeted Riemann surface. The difference equations coming from the QQ relations should then be supplemented by certain gluing conditions on the monodromies around the branch points, which define a sort of Riemann-Hilbert type problem. The Q-system difference equations and the gluing conditions together constitute the QSC which provides as an outcome the full exact spectrum of the theory.

The QSC has proven its effectiveness in a wide variety of settings and has finally made possible to harness the full power of integrability and explore highly nontrivial features of the spectrum. We give a summary of the results coming from the QSC in section \ref{sec:high}. This includes analytic calculations to 10+ loops at weak coupling, numerics with huge precision, some all-loop analytic results and studies of extreme regimes such as complexified spin or the BFKL limit. 

Very recently, a new set of applications of the QSC has started to be explored, namely the computation of correlation functions in $\cN=4$ SYM. Despite impressive progress of \cite{Basso:2015zoa} there still does not exist a satisfactory and efficient framework for computing the correlators, one major problem being related to finite-size corrections which for the spectrum are very successfully incorporated in the QSC. It has been long believed that the QSC should play a role in the context of correlators due to its close relation with Sklyanin's separation of variables (SoV) \cite{Sklyanin:1991ss,Sklyanin:1995bm}. We are finally starting to see these links being made quantitative \cite{Cavaglia:2018lxi,Giombi:2018qox,Giombi:2018hsx} and we outline these recent developments in the final parts of this review.

Given its importance in the field of AdS/CFT integrability, there exist several reviews of the QSC construction. We would like to highlight in particular the detailed and very pedagogical review \cite{Gromov:2017blm} as well as a more recent article \cite{Kazakov:2018hrh} (see also \cite{Marboe:2017zdv}). At the same time, the latest developments of the past 1-2 years are necessarily not reflected there. In the present review we aimed to keep the exposition relatively brief yet accessible, and to discuss in some detail the emerging applications of the QSC to correlators which are being actively developed.

We will not present the original derivation of the QSC from other approaches such as TBA \cite{Gromov:2014caa}, since it is rather technical and in any case applies only for some subsets of states (for which the TBA is well understood). Instead we will present the final result and describe its algebraic structure, starting from simpler spin chains as a motivating example. 

The structure of the text is as follows. We start in section \ref{sec:Qs} by describing the universal QQ relations which serve as a basis for the QSC. We discuss how these relations are formulated for the basic example of rational spin chains. We pedagogically discuss the simplest $gl(2)$ case before proceeding to generalizations for $gl(N)$ and supersymmetric $gl(M|N)$ models. Then in section \ref{sec:qsc} we present the QSC itself and outline the major results obtained using it.

Finally in section \ref{sec:correl} we describe some links between the QSC and correlation functions, in the context of cusped Wilson lines in the ladders limit for which we describe the implementation of the QSC as well. We also describe a new point of view on various components of the QSC which is starting to emerge in the framework of Sklyanin's separation of variables. In particular, we show that the gluing conditions, which are a key part of the QSC, can be interpreted as a normalizability condition for the Q-functions with respect to a simple norm.

\section{Q-functions and QQ-relations}
\label{sec:Qs}

In this section we discuss the basic functional equations underlying the QSC, which are known as QQ-relations and can be formulated for a wide variety of integrable models. We will start with simplest $gl(2)$ spin chains, and then move on to $gl(N)$ and the supersymmetric case. We will follow the notation and conventions of \cite{Kazakov:2015efa,Gromov:2014caa} and refer the reader to these papers for a more technical discussion as well as additional references.

\subsection{The $gl(2)$ case and Heisenberg spin chain}
\par\noindent
\textbf{The basics.}\ \ \ \  In order to illustrate the QQ-relations, let us start with the simplest example of the usual XXX Heisenberg spin chain, which is based on $gl(2)$ symmetry. We will discuss this case pedagogically and in detail, since it serves as a basis for all generalizations presented later.

For a spin chain with a fundamental representation (on the space $\mC^2$ with basis $\ket{\uparrow}$ and $\ket{\downarrow}$) at each site, the Hamiltonian is a sum of permutation operators on adjacent sites,
\beq
	H=\sum_{j=1}^L (1-P_{j,j+1}) \ ,
\eeq
where we assume periodic boundary conditions so $P_{L,L+1}\equiv P_{L,1}$. Its spectrum is described by Bethe equations\footnote{There are numerous reviews on Bethe ansatz, see e.g. \cite{Faddeev:1996iy,Volin:2010cq,HubbardBook,Levkovich-Maslyuk:2016kfv}.}
\beq
\label{baesu2}
   \(\frac{u_{1,j}+i/2}{u_{1,j}-i/2}\)^L=-\prod_{k=1}^K\frac{u_{1,j}-u_{1,k}+i}{u_{1,j}-u_{1,k}-i}\;,\;\ \ \ j=1,\dots,K\;,
\eeq
where $L$ is the number of sites and $K$ is the number of excitations. We labelled the roots as $u_{1,j}$ instead of $u_j$ for further convenience. The excitations correspond to flipped spins with respect to the reference (pseudovacuum) state
\beq
	\ket{0}=\ket{\uparrow\uparrow\dots \uparrow}
\eeq
which is clearly itself an eigenstate. The energies are then given by
\beq
\label{ebae}
	E=\sum_{j=1}^K\frac{1}{u_{1,j}^2+1/4} \ .
\eeq

We will focus on an alternative way to describe the solution, in which we can avoid using Bethe roots directly and instead encode them in the polynomial known as the Baxter Q-function
\beq
	Q_1(u)=\prod_{j=1}^K(u-u_{1,j}) \ .
\eeq
The energy formula \eq{ebae} is easy to rewrite in terms of this function,
\beq
	E=i\left.{\d_u}\log\frac{Q_1(u+i/2)}{Q_1(u-i/2)}\right|_{u=0}
\eeq
Importantly, rather than using Bethe equations, we can equivalently fix the Q-function from the difference equation
\beq
\label{qqul}
	Q_1^+Q_2^--Q_1^-Q_2^+=u^L
\eeq
once we require that both $Q_1$ and $Q_2$ are polynomials. This equation serves as the simplest example of a QQ relation. Here we used the notation
\beq
	f^{[n]} \equiv f(u + i n/2 ), \ \ f^\pm\equiv f(u\pm i/2) \;
\eeq
One can easily see that the Bethe equations follow from \eq{qqul}. Namely, if we shift in this equation $u\to u\pm i/2$ we get two equations
\beq
	Q_1^{++}Q_2 - Q_1Q_2^{++}=(u+i/2)^L \ , \ \ 	Q_1Q_2^{--} - Q_1^{--}Q_2=(u-i/2)^L
\eeq
Evaluating both of them on a root $u=u_{1,j}$ of $Q_1$, we find that one term in the l.h.s. vanishes in each of the equations. Then taking the ratio of the results we find precisely the Bethe equations \eq{baesu2}.

Similarly to roots of $Q_1$, the roots of the polynomial $Q_2$ appearing in \eq{qqul} also have a physical meaning and are known as the \textit{dual} Bethe roots (see e.g. \cite{Bazhanov:1996dr,Pronko:1999gh}). From \eq{qqul} it follows that the degree of $Q_2$ is $L-K+1$ so we have
\beq
	Q_2={\rm const}\times\prod_{j=1}^{L-K+1}(u-u_{2,j})
\eeq
While the usual Bethe roots correspond to excitations over the vacuum with all spins up, one can alternatively view the same excited state by starting from another reference state
\beq
	\ket{0'}=\ket{\downarrow\downarrow\dots \downarrow}
\eeq
where all spins are down. In this picture the spins up are treated as excitations and roughly speaking correspond
%\footnote{This statement is qualitative and strictly speaking is not always true. It becomes precisely true for the spin chain with a diagonal twist (described briefly below) that removes degeneracies, where the Q-functions have the form \eq{qexp}.}
 to the dual roots $u_{2,j}$. It follows from \eq{qqul} that these roots satisfy exactly the same Bethe equations \eq{baesu2}. One can verify that the two sets of roots $\{u_{1,j}\}$ and $\{u_{2,j}\}$ give the same energy eigenvalue via \eq{ebae}.
%\footnote{Depending on the construction used, the eigenstates obtained from these two sets of Bethe roots may differ, but they will lie in the same multiplet of the global $gl(2)$ symmetry of the model and have the same energy eigenvalue.}.

%In this sense the two sets of roots are on equal footing and we have chosen to label them uniformly as $u_{a,j}$ (with $a=1$ or $a=2$).

Let us also highlight the fact that the degrees of the polynomials $Q_1,Q_2$ encode the number of flipped spins $K$, or in other words the $U(1)$ charge of the state,
\beq
\label{gl2as}
	Q_1\sim \ u^K, \ Q_2 \sim u^{L-K+1}\ , \ \ u\to\infty
\eeq

The interpretation of dual roots as describing excitations over a different vacuum is qualitative and strictly speaking is not always true. It does become precisely true for the spin chain with a diagonal twist (described briefly below) that removes degeneracies, where the Q-functions have the form \eq{qexp}. For the undeformed spin chain, it is well known that in order to describe all eigenvalues one should only\footnote{We leave aside the issue of some pathological solutions involving roots at e.g. $\pm i/2$, see e.g. \cite{Hao:2013jqa,Granet:2019knz} for more details and references on this subtle point. In fact the Q-system approach is often more efficient than the Bethe equations at capturing these 'singular' solutions.} consider solutions of Bethe equations \eq{baesu2} with $K\leq L/2$. For each such solution, the corresponding set of the dual roots $\{u_{2,j}\}$ will have more than $L/2$ elements and thus one does not usually consider these solutions. Sometimes they may have some strange features, e.g. not be fixed uniquely\footnote{The reason for this is that when the degree of $Q_1$ as a polynomial in $u$ is smaller than that of $Q_2$, one can replace $Q_2$ by a linear combination $Q_2\to Q_2+{\rm const}\cdot Q_1$ which will shift the dual roots but still solve the QQ relation \eq{qqul} and preserve the asymptotics \eq{gl2as}.}. Nevertheless these dual roots play an important role in understanding algebraic structures behind the Bethe equations.

One can also easily make a connection to the usual Baxter equation. Let us consider the determinant
\beq
	D(u)=\begin{vmatrix}
	Q^{++} & Q & Q^{--} \\
	Q_1^{++} & Q_1 & Q_1^{--} \\
	Q_2^{++} & Q_2 & Q_2^{--}
	\end{vmatrix}
\eeq
which clearly vanishes when $Q$ is any linear combination of $Q_1$ and $Q_2$. Expanding it in the first row and using \eq{qqul}, we get
\beq
\label{baxgl2}
	Q^{++}(u-i/2)^L + Q^{--}(u+i/2)^L = TQ
\eeq
where we defined the polynomial
\beq
	T=\begin{vmatrix}
	Q_1^{++} & Q_1^{--} \\
	Q_2^{++} & Q_2^{--}
	\end{vmatrix}
\eeq
which should be identified with the transfer matrix eigenvalue. We see that \eq{baxgl2} is precisely the usual Baxter equations. As a second order difference equation, it has two solutions which will be $Q_1$ and $Q_2$ once we impose that they (as well as $T$) are polynomials.

\par\noindent
\textbf{Canonical form of the QQ relation.}\ \ \ \  For what follows it is instructive to rewrite \eq{qqul} in the canonical form of a QQ relation by introducing two more Q-functions,
\beq
	Q_\emptyset=u^L, \ \ Q_{12}=1
\eeq
Then \eq{qqul} takes the standard bilinear form
\beq
\label{qqgl2}
	Q_1^+Q_2^--Q_1^-Q_2^+=Q_\emptyset Q_{12}
\eeq
As a result, for the $gl(2)$ spin chain we have four polynomial Q-functions that we can label as $Q_A$ with $A$ being a multi-index consisting of up to two distinct elements from the set $\{1,2\}$. In general one can always set $Q_{12}=1$ by performing a gauge transformation on the Q-functions (see e.g. \cite{Kazakov:2015efa}). When this is done, the function $Q_\emptyset$ is determined by the particular model we are considering. For example, generalizing to a spin chain with inhomogeneities $\theta_1,\dots,\theta_L$ simply amounts to taking
\beq
	Q_\emptyset=\prod_{j=1}^L(u-\theta_j)
\eeq
and imposing the same QQ relation \eq{qqgl2} which will lead to the standard Bethe equations once we require $Q_1,Q_2$ to be polynomials. Having done that, one can also introduce twisted boundary conditions with a $e^{i\phi}$ twist by allowing $Q_i$ to have exponential behavior,
\beq
\label{qexp}
	Q_{1}={\rm const}\times e^{u\phi}\times\[\text{polynomial}\]\ , \ Q_{2}={\rm const}\times e^{-u\phi}\times\[\text{polynomial}\]
\eeq
which will give the expected Bethe equations
\beq
    \prod_{n=1}^L\frac{u_j-\theta_n+i/2}{u_j-\theta_n-i/2}=-e^{2i\phi}\prod_{k=1}^K\frac{u_j-u_k+i}{u_j-u_k-i}\;,\;\ \ \ j=1,\dots,K\;.
\eeq
As another example, taking
\beq
	Q_\emptyset=1/u^L
\eeq
would correspond to a spin chain with an infinite-dimensional $s=-1/2$ representation at each site.

Thus we see that the QQ relation \eq{qqgl2} is very general and can describe various versions of the $gl(2)$ spin chain. Let us summarize several key points:
\begin{itemize}
	\item The form of the QQ relations is determined by the symmetry of the model. This is a somewhat trivial statement here for $gl(2)$ but we will soon see how it works for any $gl(N)$.
	\item The choice of the boundary Q-function $Q_\es$ contains information about the specific model we are considering.
	\item Large $u$ asymptotics of the Q-functions determines the charges of the state (see \eq{gl2as}).
	\item The Q-functions are fixed by imposing the QQ relations, the large $u$ behavior and, very importantly, the analytic properties (for simplest $gl(2)$ spin chains this is just polynomiality).
\end{itemize}
All of these properties generalize to the Quantum Spectral Curve description of $\cN=4$ SYM.

%Thus we see that the QQ relation \eq{qqgl2} is very general and can describe various versions of the $gl(2)$ spin chain, depending on how we choose the `boundary' Q-function $Q_\emptyset$ and what properties we impose on the solutions $Q_1,Q_2$.

In the next section we discuss how this picture can be extended to the $gl(N)$ case.

\subsection{Generalization to $gl(N)$}

The standard rational $gl(N)$ spin chain can be solved by the Bethe ansatz methods similarly to the $gl(2)$ case. In this case the Bethe equations have a \textit{nested} structure and involve several types of Bethe roots \cite{SutherlandN1,Kulish:1983rd}. While the Bethe equations have a rather concise form, the direct analog of the Baxter equation is an $N$-th order diffeence equation that is technically somewhat harder to write down. The Q-system, on the other hand, has essentially the same form for any $gl(N)$ as we will see soon.

Let us first discuss the $gl(3)$ spin chain as an example. The Bethe equations contain the momentum-carrying roots $u_{1,j}$ and the auxiliary roots $u_{12,j}$, and read for the inhomogeneous chain
\beqa\la{Bethe31}
    \prod_{n=1}^L\frac{u_{1,j}-\theta_n+i/2}{u_{1,j}-\theta_n-i/2}&=&\prod_{k\neq j}^{K_1}\frac{u_{1,j}-u_{1,k}+i}{u_{1,j}-u_{1,k}-i}
    \prod_{l=1}^{K_{12}}\frac{u_{1,j}-u_{12,l}-i/2}{u_{1,j}-u_{12,l}+i/2}\ , \\ \la{Bethe32}
    1&=&\prod_{k\neq j}^{K_{12}}\frac{u_{12,j}-u_{12,k}+i}{u_{12,j}-u_{12,k}-i}\prod_{l=1}^{K_1}\frac{u_{12,j}-u_{1,l}-i/2}{u_{12,j}-u_{1,l}+i/2}\;,
\eeqa
The roots $u_{12,j}$ can be understood as excitations propagating on an auxiliary $gl(2)$ chain of length $K_1$ for which inhomogeneities are the momentum-carrying roots $u_{1,j}$.

Like in the $gl(2)$ case, we can rewrite these equations in terms of various sets of dual roots. Now at each site of the chain we have a $\mC^3$ space so there are three equivalent choices of the pseudovacuum state, and accordingly for each state we have three sets of momentum-carrying roots that we can label as $\{u_{1,j}\}, \;\{u_{2,j}\}$ and $\{u_{3,j}\}$. Thus we can introduce the corresponding Q-functions
\beq
	Q_a=\prod_{j=1}^{K_a}(u-u_{a,j})\ , \ \ \ a=1,2,3
\eeq
We can additionally choose which of the two pseudovacua to use for the auxiliary $gl(2)$ chain. Some of these choices are not completely independent, however. As a result, we can encompass all possible duality transformations by introducing 3 more\footnote{Another reason why Q-functions with more than one index are important is that they appear to play a key role in the separation of variables for higher rank models, as observed recently in \cite{Cavaglia:2019pow}.} Q-functions $Q_{12}, Q_{13}$ and $Q_{23}$,
\beq
	Q_{A}=\prod_{j=1}^{K_A}(u-u_{A,j})\ , \ \ \ A=(12),\;(13),\;(23)
\eeq
Instead of writing all possible dual sets of Bethe equations, we can simply impose on these Q-functions the canonical QQ relations which have the universal form
\beq
\label{qqgln}
	Q_AQ_{Aab}=Q_{Aa}^+Q_{Ab}^- - Q_{Aa}^-Q_{Ab}^+ 
\eeq
similarly to \eq{qqgl2} that we discussed for the $gl(2)$ case. Here $A$ is a multi-index taking values in the set $\{1,2,3\}$. We also fix
\beq
\label{qes3}
	Q_\emptyset=\prod_{n=1}^L(u-\theta_n) \ , \ \ Q_{123}=1
\eeq
and we require the Q-functions to be antisymmetric in their indices.
Then imposing the QQ relations and requiring polynomiality of the Q-functions provides a discrete set of solutions corresponding to states of the spin chain. One can derive the Bethe equations from these QQ relations just like we did for the $gl(2)$ case above.

The same approach directly generalizes to the case of any $gl(N)$. Now we have $2^N$ different Q-functions $Q_A$ labeled by a multi-index $A$ from the set $\{1,2,\dots,N\}$. Again the Q-functions are antisymmetric in their indices, and each Q-function encodes a particular set of Bethe roots on the corresponding level of nesting.

\begin{figure}[ht]
\begin{center}
\includegraphics[scale=0.45]{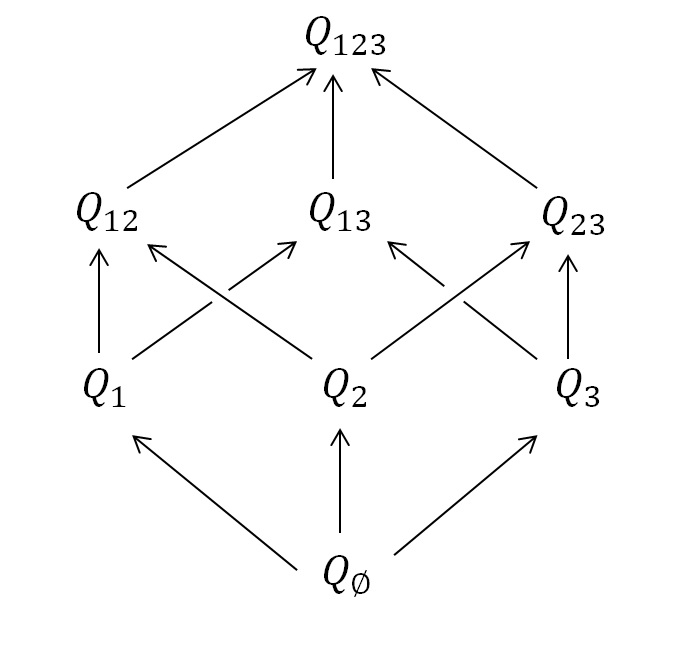}
\end{center}
\caption{\label{fig:hasse1} The Hasse diagram for a $gl(3)$ spin chain. Each vertex corresponds to a Q-function, and each face with 4 vertices corresponds to a QQ relation \eq{qqgln}. Every path going from $Q_{\emptyset}$ to $Q_{123}$ (with 1 index added at each step) corresponds to a choice of Bethe equations. }
\end{figure}

A useful way to view the QQ relations is to consider the \textit{Hasse diagram}, shown on figure \ref{fig:hasse1} for the $gl(3)$ example. Each node in this diagram corresponds to a Q-function, while each face with 4 vertices corresponds to one of the bilinear QQ relations \eq{qqgln}. In order to write the Bethe equations (with a particular choice of which of the dual roots we use at each level), one should choose a path on this diagram going from $Q_\emptyset$ to $Q_{123}$ in which successive nodes correspond to Q-functions with increasing number of indices. Then from the QQ relations we find the set of Bethe equations
\beqa
	\frac{Q_\es^+}{Q_\es^-}&=&\left.-\frac{Q_a^{++}}{Q_a^{--}}\frac{Q_{ab}^{-}}{Q_{ab}^{+}}\right|_{u=u_{a,n}} \\
	1&=&\left.-\frac{Q_{ab}^{++}}{Q_{ab}^{--}}\frac{Q_{a}^{-}}{Q_{a}^{+}}\right|_{u=u_{ab,n}}
\eeqa
(with $Q_\es$ given by \eq{qes3}) which involve the roots of Q-functions $Q_a,Q_{ab}$ along the path.

Let us mention that in the classical limit going from one dual set of Bethe roots to another corresponds to reordering the sheets of the classical spectral curve, see \cite{Gromov:2007ky} for a detailed discussion.

For what follows it will be important to introduce the \textit{Hodge dual} Q-functions. They carry upper instead of lower indices and are defined as
\beq
\label{hodge}
	Q^A=\epsilon^{A'A}Q_{A'} \ , \ \ A'=\{1,\dots,N\}\backslash A
\eeq
where we emphasize that there is \textit{no summation} over the repeated indices. Here $\epsilon$ is the fully antisymmetric Levi-Civita tensor. In other words, these new Q-functions are obtained from the original ones by a relabeling and a possible sign change. The main property of $Q^A$ is that they satisfy exactly the same QQ relations \eq{qqgln} as before, provided $Q_A$ satisfy them. Thus, for every solution of the QQ relations we can construct a new one using \eq{hodge}.

We would like to note the similarity between the bilinear QQ relation and the Pl\"ucker relations for coordinates on a Grassmannian. For a somewhat different geometrical interpretation of the Q-system, see \cite{Kazakov:2015efa}.

In the next section we discuss how the construction generalizes to supersymmetric spin chains.

\subsection{Generalization to $gl(M|N)$}

The spin chains based on the superalgebra $gl(M|N)$ can be described by QQ relations just as well as their bosonic counterparts (see e.g. \cite{Kazakov:2007fy,Kazakov:2010iu}). This generalization is important for us since it is the basis for the QSC of $\cN=4$ SYM, where the global symmetry of the model leads to a $gl(4|4)$ Q-system.

The Q-functions for $gl(M|N)$ are denoted as $Q_{A|I}$ where $A$ is a multi-index from the set $\{1,2,\dots,M\}$ and $I$ is a multi-index from the set $\{1,2,\dots,N\}$, so that we have $2^{M+N}$ Q-functions as a result. We will refer to the indices in $A$ as bosonic and those in $I$ as fermionic.

Somewhat surprisingly, the supersymmetric Q-systems can be obtained from the bosonic case simply by relabelling the Q's. Namely, we have\footnote{Ultimately this procedure has intriguing relations with links between representation theory of usual and of graded Lie algebras, see \cite{Kazakov:2015efa} and references therein.},
\beq
\label{bfmap}
	Q_{A|I}=\epsilon^{{\bar I}I}q_{A\{\bar I + M\}} \ , \ \ \ \ \bar I=\{1,\dots,N\}\backslash I
\eeq
with no summation over repeated indices. Here $q_C$ solve a $gl(M+N)$ bosonic Q-system, and the notation $\{\bar I + M\}$ means we add $M$ to each element of $\bar I$. As before, in \eq{bfmap} we have no summation over the repeated indices. We refer the reader to \cite{Gromov:2010km} for details on this procedure, which essentially corresponds to a ninety-degrees rotation of the Hasse diagram. Nevertheless, it will be convenient to use the notation $Q_{A|I}$ specifically adapted to the graded case. Then, instead of just one QQ relation \eq{qqgln} for the bosonic case, we have three types of relations,
\beqa
\label{qqs1}
	Q_{A|I}Q_{Aab|I}&=&Q_{Aa|I}^+Q_{Ab|I}^- - Q_{Aa|I}^-Q_{Ab|I}^+ \\
	\label{qqs2}
	Q_{A|I}Q_{A|Iij}&=&Q_{A|Ii}^+Q_{A|Ij}^- - Q_{A|Ii}^-Q_{A|Ij}^+ \\
	\label{qqs3}
	Q_{Aa|I}Q_{A|Ii}&=&Q_{Aa|Ii}^+Q_{A|I}^- - Q_{Aa|Ii}^-Q_{A|I}^+
\eeqa
The first two of these exchange indices of the same type (either bosonic or fermionic) and are known as bosonic QQ relations. The third one exchanges indices of different gradings and is called the fermionic QQ relation. As before, these relations correspond to various duality transformations on the Bethe roots in Bethe equations. 

Just as before, for the spin chain with a fundamental representation at each site we require all Q's to be polynomial and impose boundary conditions
\beq
	Q_{\emptyset|\emptyset}=u^L, \ \ Q_{\bar\emptyset|\bar\emptyset}=1
\eeq
Then, choosing a path on the Hasse diagram going from $Q_{\es|\es}$ to $Q_{\bes|\bes}$, where at each step we add a new index, we can derive a set of Bethe equations in a particular grading, similarly to the bosonic case. Due to the slightly different structure of the bosonic and fermionic QQ relations (\eq{qqs1} and \eq{qqs2}, \eq{qqs3}), we will find Bethe equations of two types (with bosonic or fermionic Bethe roots), as expected.

For the supersymmetric Q-system we can define the Hodge dual Q-functions similarly to the bosonic case. Instead of \eq{hodge} we now introduce
\beq
\label{hodgesu}
	Q^{A|I}=(-1)^{|A'||I|}\epsilon^{A'A}\epsilon^{I'I}Q_{A'|I'} \ , \ \ \ A'=\{1,\dots,M\}\backslash A \ , \ \ I'=\{1,\dots,N\}\backslash I
\eeq
(again without summation over repeated indices). One can show that they will satisfy exactly the same QQ relations as the original set of Q-functions. In the QSC for $\cN=4$ SYM the Hodge dual Q's play an important role as we will soon discuss.

\section{The QSC for $\cN=4$ super Yang-Mills}
\label{sec:qsc}

In this section we describe the Quantum Spectral Curve construction for $\cN=4$ SYM. 

For compact spin chains discussed above, the Q-functions were fixed by two requirements -- first, they should satisfy the QQ relations, and second, they should be polynomials. We will first describe the QQ relations underlying the QSC for $\cN=4$ SYM, and then discuss the analytic properties one should impose instead of polynomiality.

\subsection{The Q-system}

The QSC is based on the Q-system corresponding to the global $psu(2,2|4)$ symmetry of the theory, which is a particular real form of $gl(4|4)$ supplemented with some other restrictions such as unimodularity. As a result, the QQ relations have the same universal form \eq{qqs1}, \eq{qqs2}, \eq{qqs3} as for the $gl(4|4)$ spin chain and thus involve $2^8=256$ Q-functions, labelled as $Q_{A|I}$ where $A$ and $I$ are multi-indices from the set $\{1,2,3,4\}$.
One important difference is in the choice of boundary Q-functions $Q_{\es|\es}$ and $Q_{\bes|\bes}$ which as we discussed typically vary depending on the model. While for spin chains one of them is simply $1$ and the other one is a fixed polynomial such as $u^L$, here we require
\beq
\label{unim}
	Q_{\es|\es}=1, \ \ Q_{\bes|\bes}=1
\eeq
This requirement may be viewed as the counterpart of the unimodularity conditions for $psu(2,2|4)$, see \cite{Gromov:2014caa} for more details.

Let us discuss several useful consequences of the QQ relations. It is convenient to introduce special notation for some of the Q-functions and their Hodge duals, namely
\beq
	\bP_a=Q_{a|\es} \  , \ \bP^a=Q^{a|\es} \ ,  \ \bQ_i=Q_{\es|i} \ , \ \bQ^i=Q^{\es|i}
\eeq
The $\bP$-functions carry indices corresponding to the $SU(4)$ R-symmetry subgroup of the global symmetry, and accordingly are related (roughly speaking) to dynamics of the dual string on the $S^5$ part of the ${\rm AdS}_5\times S^5$ background. Similarly, the $\bQ$-functions have indices corresponding to the $SU(2,2)$ conformal group and roughly speaking carry information about the dynamics on ${\rm AdS}_5$.

One may choose some subset of these functions as the `basis' Q-functions, and then other Q's are fixed in terms of them (at least in principle) by solving the QQ relations. One frequently used choice is the set of 4+4 functions $\bP_a$ and $\bP^a$.

Let us mention that in left-right symmetric sectors such as the $sl(2)$ sector, the whole construction simplifies as we have a simple relation for the Q-functions with upper and lower indices, namely $\bP^a=\chi^{ab}\bP_b$ and $\bQ^i=\chi^{ij}\bQ_j$ with the only non-zero entries of $\chi^{ab}$ being $\chi^{14}=-\chi^{23}=\chi^{32}=-\chi^{41}=-1\ $.

Another important subset of Q-functions are the $Q_{a|i}$. They satisfy the QQ relation
\beq
\label{QPQ}
	Q_{a|i}^+-Q_{a|i}^-=\bP_a\bQ_i
\eeq
which is just the equation \eq{qqs1}. In turn one can prove another very useful relation, namely that \cite{Gromov:2014caa} 
\beq
\label{QisPQ}
	\bQ_i=-\bP^aQ_{a|i}^+
\eeq
Due to the choice \eq{unim} we have several extra nice properties that follow from the QQ relations, as discussed in \cite{Gromov:2014caa}, for example
\beq
	Q^{a|i}Q_{a|j}=-\delta^i_j \ , \ \ Q^{a|i}Q_{b|i}=-\delta^a_b \ ,
\eeq
and 
\beq
	\bP^a\bP_a=0, \ \ \bQ^i\bQ_i=0
\eeq

One reason for the importance of $Q_{a|i}$ is that one can express any Q-function explicitly (i.e. without any infinite sums) in terms of them together with $\bP_a, \bQ_i$, for example\footnote{From \eq{QPQ} we see that $Q_{a|i}$ itself is fixed in terms of $\bP_a$ and $\bQ_i$, but the solution would involve an infinite sum, e.g. $Q_{a|i}=-\sum_{n=1}^\infty \bP_a^{[2n-1]}\bQ_i^{[2n-1]}+{\cal P}$ where ${\cal P}$ is an $i$-periodic function.} 
\beq
	Q_{ab|\es} = \begin{vmatrix}
	Q_{a|\es}^+ & Q_{a|\es}^- \\
	Q_{b|\es}^+ & Q_{b|\es}^- \\
	\end{vmatrix}
	\ , \ \ \ Q_{ab|ij}=\begin{vmatrix}
	Q_{a|i} & Q_{b|i} \\
	Q_{a|j} & Q_{b|j}
	\end{vmatrix}
\eeq
A wide variety of similar formulas exists also for standard spin chains.

In applications of the QSC it is often useful to derive an equations for the $\bQ$'s in terms of $\bP$'s. This equation, which follows from the QQ relations above, has the form \cite{Alfimov:2014bwa}
\beqa\la{bax5}
\bQ^{[+4]}_iD_0
&-&
\bQ^{[+2]}
\[
D_1-\bP_a^{[+2]}\bP^{a[+4]}D_0
\]
+
\bQ
\[
D_2-\bP_a\bP^{a[+2]}D_1+
\bP_a\bP^{a[+4]}D_0
\]\\
&-&
\bQ^{[-2]}
\[
\bar D_1+\bP_a^{[-2]}\bP^{a[-4]}\bar D_0
\]
+\bQ^{[-4]}\bar D_0
=0\nn \ ,
\eeqa
where $D_n, \bar D_n$ are some determinants constructed from $\bP_a$ which we give explicitly in Appendix A.

\subsection{Large $u$ asymptotics}

As we already saw for the spin chains, the large $u$ asymptotics of the Q-functions encodes the global charges of the state. For $\cN=4$ SYM these are the Cartan charges of the bosonic $SU(2,2)\times SU(4)$ global symmetry, namely $(\Delta,S_1,S_2)$ for the conformal symmetry $SU(2,2)$ and $(J_1,J_2,J_3)$ for the $SU(4)$ R-symmetry. The charges $S_1,S_2$ are the spins in ${\rm AdS_5}$, while $J_1,J_2,J_3$ are the three angular momenta on $S^5$. Of course, $\Delta$ is particularly important since it is the conformal dimension of the state, which will be fixed by the QSC as a function of the coupling $\lambda$. 

Explicitly, the asymptotics read
\beq
	\bP_a\sim A_au^{-\tilde M_a}, \ \ \bP^a\sim A^au^{\tilde M_a-1}\ , \ \ \bQ_i\sim B_iu^{\hat M_i-1} \ , \ \ \bQ^i\sim B^iu^{-\hat M_i}
\label{asymptPQ}
\eeq
where
\begin{align}
\label{relMta}
\tilde M_a=
&\left\{\frac{J_1+J_2-J_3+2}{2}
  ,\frac{J_1-J_2+J_3}{2}
   ,\frac{-J_1+J_2+J_3+2}{2}
   ,\frac{-J_1-J_2-J_3}{2}
   \right\}\ ,\\
\hat M_i=&
\left\{
\frac{\Delta -S_1-S_2+2}{2} ,
\frac{\Delta +S_1+S_2}{2}
   ,
\frac{-\Delta-S_1+S_2+2}{2} ,
\frac{-\Delta+S_1-S_2}{2} \right\}\ .
\label{M-ass}\end{align}
We see that $\bQ_i$ have non-integer asymptotics, reflecting the noncompactness of the conformal symmetry group. One can also use \eq{QPQ}, \eq{QisPQ} to show that the leading coefficients satisfy\footnote{There is a similar relation for $B_{i_0}B^{i_0}$.}
\beq
\label{AA}
	A^{a_0}A_{a_0}=i\frac{\prod_j (\tilde M_{a_0}-\hat M_j)}{\prod_b (\tilde M_{a_0}-\tilde M_b)}
\eeq
with no summation over $a_0$.

These asymptotics can be to a large extent read off from the classical limit, where the Q-functions are related to the quasimomenta of the classical spectral curve (see section \ref{sec:class}). The extra shifts by $\pm 1$ are not visible in this limit, but can be seen by a comparison with the asymptotic Bethe ansatz which can be derived from the QSC \cite{Gromov:2014caa}.

\subsection{Analyticity and gluing conditions}

\begin{figure}[ht]
\begin{center}
\includegraphics[scale=0.45]{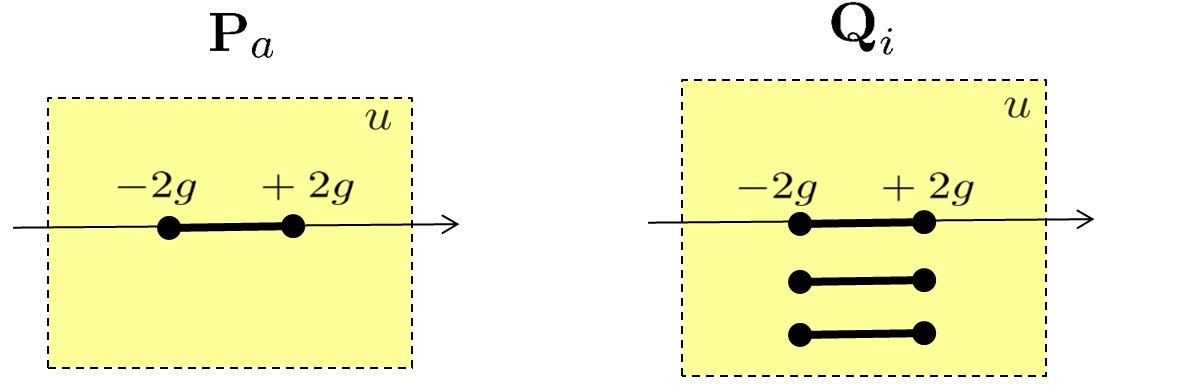}
\end{center}
\caption{\label{fig:QP1} The cut structure of the functions $\bP_a$ and $\bQ_i$. }
\end{figure}

While the QQ relations and asymptotics are largely dictated by symmetry, the most nontrivial part of the QSC are the analytic properties of the Q-functions. While in other integrable models such as spin chains the Q's are typically meromorphic or even polynomial, here they have branch cuts whose position is fixed by the 't Hooft coupling $\lambda$. The $\bP$-functions have only one cut, located at $u\in[-2g,2g]$ where
\beq
	g=\frac{\sqrt{\lambda}}{4\pi}
\eeq
Cuts of this type have of course appeared already in the asymptotic Bethe ansatz for SYM where they originate from the Zhukovsky variable $x(u)$,
\beq
\label{defx}
    x(u)=\frac{u+\sqrt{u-2g}\sqrt{u+2g}}{2g}\;\;,\;\;u=(x+1/x)g\;
\eeq
Yet other Q-functions typically have infinitely many branch points, although no other singularities are allowed in any Q-functions. It is standard to choose all the Q's to be analytic in the upper half plane, then for example $\bQ_i(u)$ and $\bQ^i(u)$ have cuts at $[-2g-in,2g-in]$ with $n=0,1,2,\dots$ (see figure \ref{fig:QP1}). Let us also note that, due to QQ relations, if we have a cut in some Q-function it will typically propagate and lead to an infinite set of cuts spaced by $i$ in other Q's.

\begin{figure}[ht]
\begin{center}
\includegraphics[scale=0.45]{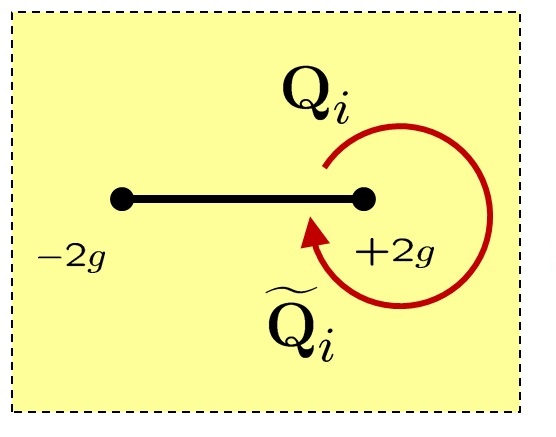}
\end{center}
\caption{\label{fig:Qtil} We denote by tilde the operation of analytic continuation around the branch point at $u=2g$. The gluing conditions relate $\tilde \bQ_i$ with $\bar \bQ^j$.}
\end{figure}

In order to fix the Q-functions and thus the dimension $\Delta$, we will impose \textit{gluing conditions} which involve monodromy around the cut endpoints. Let us denote by tilde the operation of going around the branch point at $u=2g$, so that if we analytically continue some Q-function $Q(u)$ through the cut we get a new function $\tilde Q(u)$, see figure \ref{fig:Qtil}. For example, applying this operation to $x(u)$ gives $1/x(u)$. The gluing conditions which we should impose then have the form
\beq
\label{glc}
	\tilde\bQ_1=\alpha_1\bar\bQ^2\ , \ \tilde\bQ_2=\alpha_2\bar\bQ^1,
	\ \tilde\bQ_3=\alpha_3\bar\bQ^4, \ \tilde\bQ_4=\alpha_4\bar\bQ^3
\eeq
where $\alpha_i$ are some constants and bar denotes complex conjugation. Curiously, these relations mean that $\tilde \bQ_i$ are analytic in the lower half-plane, while the original $\bQ_i$ were analytic in the upper half plane.

This form of the gluing conditions was proposed in \cite{Gromov:2015vua} and was not historically the first one. The original formulation of the QSC \cite{Gromov:2013pga} involved a different version of the monodromy constraints, utilizing extra auxiliary functions known as $\mu_{ab}$ and $\omega_{ij}$. Later it was realized in \cite{Gromov:2015vua} that they can be eliminated, leading to the highly compact equations \eq{glc} which have proven very efficient in various settings.

The gluing conditions \eq{glc} are a crucial part of the QSC. Together with the QQ relations they fix the Q's completely (up to some global symmetries) and determine $\Delta$ as a function of the coupling $\lambda$. Notice that the coupling appears in the formulation only through the position of the branch points.

Now we can state a bit more precisely how one can solve the QSC equations in practice in various situations (e.g. numerically or perturbatively). Typically we proceed in several steps:
\begin{itemize}

\item We start by introducing an efficient parametrization for the $\bP$-functions. Since they have only one cut, we can resolve it by switching to $x(u)$ and write them as Laurent series in $x$,
\beq\label{param}
\bP_a(u)=\sum_{n=\tilde M_a}^\infty \frac{c_{a,n}}{x^n(u)}\; \ \ , \ \ 
\bP^a(u)=\sum_{n=-\tilde M_a+1}^\infty \frac{c^{a}_n}{x^n(u)}
\eeq
where $c$'s are some $u$-independent coefficients that are the main unknowns in the problem and encode all the nontrivial data about the state. In particular from them one can compute $\Delta$ using e.g. \eq{AA}.

\item Next we should solve the Baxter-like 4th order equation \eq{bax5} and a similar equation for $\bQ^i$, which indirectly determines $\bQ_i$ and $\bQ^i$ in terms of the $c_{a,n}$ and $c^a_n$. The 4 solutions of this equation are identified by their asymptotics \eq{asymptPQ} which are fixed by the global charges of the state.

\item Finally we should impose the gluing conditions \eq{glc} which will fix the values of the coefficients $c_{a,n}$ and $c^a_n$, thus determining the conformal dimension.

\end{itemize}

We briefly discuss an example of solving the QSC equations in section \ref{sec:correl}, for the QSC describing a slightly different observable (cusped Wilson line). Several pedagogical examples of calculations with QSC can be found in \cite{Gromov:2017blm}.

\subsection{Classical limit and spectral curve}
\label{sec:class}

Naturally, in the classical limit the QSC encodes the spectral curve originating from classical integrability of the superstring sigma model. Namely, one can show that (see \cite{Gromov:2014caa} for details) in the classical limit when $\Delta$ and other charges scale as $\sim\sqrt{\lambda}$ with $\lambda\to\infty$, the basic Q-functions take the form
\beq
	\bP_a\sim \exp\(-\int^u p_a(v)dv\) \ , \ \ \bP^a\sim \exp\(+\int^u p_a(v)dv\)
\eeq
\beq
	\bQ_i\sim \exp\(+\int^u q_i(v)dv\) \ , \ \ \bQ^i\sim \exp\(-\int^u q_i(v)dv\)
\eeq
where $p_a$ and $q_i$ are the 4+4 quasimomenta of the classical spectral curve corresponding to $S^5$ and ${\rm AdS}_5$ respectively. 

In fact, in general the difference equations such as QQ relations or the Baxter equations can often be viewed as a sort of quantization of the classical spectral curve. Some motivation for this comes from the classical picture of separation of variables, we refer to \cite{BabelonInt,Chervov:2006xk} for more details and references. Very roughly speaking, at the classical level the eigenvalues $\lambda$ of the classical monodromy matrix (evaluated at certain dynamical points on the curve) play the role of (exponents of) Poisson-conjugate variables to the spectral parameter. After quantization then $\lambda$ becomes a shift operator with respect to $u$, and the classical spectral curve equation becomes a 2nd order difference equation which in fact is equivalent to the usual Baxter equation. Similarly, one expects that the QSC of $\cN=4$ SYM is a quantization of the classical curve for the string sigma model. It would be highly interesting to make this statement more precise or even prove it.

\subsection{Highlights of QSC-based results}
\label{sec:high}

Here we summarize the main results obtained with the use of the QSC.

\begin{itemize}
	\item From the start it was evident that the QSC can finally allow one to compute the spectrum to very high loop order at weak coupling. Initially, 11 loops were reached, and can be computed on a usual laptop \cite{Marboe:2014gma}. Even higher orders were computed later in some cases using an efficient algorithm which is available in Mathematica form, with the result not restricted to any particular sector and essentially limited only by computing power \cite{Marboe:2017dmb,Marboe:2018ugv}. It also makes it possible to test or even prove some conjectures regarding the class of numbers (multiple zeta values) appearing as coefficients in the expansion. As a byproduct, a highly efficient method to solve nested Bethe equations was proposed in \cite{Marboe:2016yyn}. Moreover, one can compute the dimension of some operators as a function of the spin analytically to high loop order \cite{Marboe:2014sya,Marboe:2016igj}. 
	
	\item For operators from $sl(2)$ sector built from scalars and covariant derivatives, of the form $\cO=\Tr(Z{\cal D}^SZ^{L-1})+\dots$, one can study the small spin expansion of the spectrum,
	\beq
		\Delta=L+S+\gamma_L^{(1)}(\lambda)S+\gamma_L^{(2)}(\lambda)S^2+\cO(S^3)
\eeq
	The first two nontrivial coefficients were computed analytically from the QSC in \cite{Gromov:2014bva} (the first was known from asymptotic Bethe ansatz \cite{Basso:2011rs}). A re-expansion procedure converts this data into a prediction for the first \textit{four} orders of the strong coupling expansion for the dimension of  operators with fixed spin \cite{Gromov:2014bva}, such as  the well-studied Konishi operator which has $S=2$.
	
	\item It took some time to design an efficient method for solving the QSC numerically, which was presented in \cite{Gromov:2015wca}. Its precision is virtually unlimited and 60+ digits have been reached. It also works for complexified spin and allows one to uncover the rich analytic structure of $\Delta$ as a function of $S$. For example one can plot the BFKL intercept, i.e. the value of $\Delta$ analytically continued to $S=0$, all the way from weak to strong coupling (reproducing and extending known predictions). Let us mention that recently the properties of $\Delta(S)$ were discussed in the context of conformal bootstrap \cite{Kravchuk:2018htv}.

\item In \cite{Gromov:2015vua} an analytic iterative solution method was proposed which works in many different situations. It was used to compute $\Delta$ in the BFKL limit when $S\to -1,g\to 0$ with $g^2/(S+1)$ fixed, confirming old results and generating new predictions. In this limit the 'most complicated' (highest transcendentality) part of the result is the same in SYM and in large $N_c$ QCD. Important initial steps towards implementing this limit for the QSC were taken in \cite{Alfimov:2014bwa}. The results on BFKL were further greatly extended in \cite{Alfimov:2018cms}.

\item The QSC has been also formulated for the Hubbard model \cite{Cavaglia:2015nta,VolinHub} and for the ABJM theory \cite{Cavaglia:2014exa,Bombardelli:2017vhk}. For ABJM it provided high-order perturbative results and very precise numerics \cite{Anselmetti:2015mda,Lee:2017mhh,Lee:2019oml,Lee:2019bgq,Bombardelli:2018bqz}, the latter confirming and extending a much older TBA calculation \cite{LevkovichMaslyuk:2011ty}. For ABJM one can again compute the small $S$ expansion analytically which led to a conjecture for the exact form of the function $h(\lambda)$ appearing in all integrability-based results for ABJM \cite{Gromov:2014eha}. That conjecture was later extended to ABJ theory \cite{Cavaglia:2016ide} (see also the recent review \cite{Drukker:2019bev} for a recent related review).

\item The QSC can be extended to describe the generalized cusp anomalous dimension controlling the divergence of a cusped Wilson line. This was done in \cite{Gromov:2015dfa}  leading to its calculation in a near-BPS limit analytically and in various regimes numerically. In the near-BPS limit one can also make a connection to localization exact results \cite{Gromov:2013qga} (see also \cite{Gromov:2012eu}). Then in \cite{Gromov:2016rrp} the limit corresponding to the quark-antiquark potential was explored in detail. Later in \cite{Cavaglia:2018lxi} the QSC was shown to naturally incorporate the previously missing integrability description of particular scalar insertions at the cusp.

\item In \cite{Harmark:2017yrv,Harmark:2018red} a modified version of the QSC was used to efficiently compute the Hagedorn temperature in $\cN=4$ SYM in various regimes. The connection to QSC was found by reformulating the problem in a TBA language. Curiously, the analytic properties of the $\bP_a$ and $\bQ_i$ are interchanged in this setup.

\item The QSC was generalized in \cite{Kazakov:2015efa} to integrable (diagonal) twist deformations of SYM, such as the $\gamma$-deformation which breaks all supersymmetries. Some weak coupling results for this case were obtained recently in \cite{Marboe:2019wyc}. The QSC is also known for the $q$-deformation \cite{Klabbers:2017vtw}.

\item A limit of large twist in the $\gamma$-deformation leads to the fishnet theory \cite{Gurdogan:2015csr}, which in the simplest case is just a theory of two complex scalars,
\beq
	\cL=\frac{N_c}{2}\Tr\[\d^\mu\Phi_1^\dagger\d_\mu\Phi_1  + \d^\mu\Phi_2^\dagger\d_\mu\Phi_2 +2\xi^2\Phi_1^\dagger\Phi_2^\dagger\Phi_1\Phi_2\]
\eeq
This is a remarkable 'baby' version of SYM which is non-unitary but still conformal (up to some subtleties) and integrable, the integrability being visible at the level of Feynman graphs.	Implementing the limit in the QSC led to a plethora of results for this model, allowing one to make use of the powerful methods developed for SYM \cite{Gromov:2017cja} (see also the review \cite{Kazakov:2018hrh}). A very similar limit in the QSC for the quark-antiquark potential was studied  earlier in \cite{Gromov:2016rrp}.
	
\item Since recently, the expected links between QSC and 3-pt correlators are starting to be made quantitative. In \cite{Giombi:2018hsx,Giombi:2018qox} such results were found in the near-BPS limit studied using localization and utilizing some QSC results of \cite{Gromov:2013qga}. In \cite{Cavaglia:2018lxi} huge simplifications for correlators were observed in the very different ladders limit for Wilson lines, highly similar to the fishnet theory. We review some of these results in the next section.
	
\end{itemize}

\section{QSC for cusps and links with correlation functions}
\label{sec:correl}

In this section we describe a generalization of the QSC for the cusped Wilson line. The reason this setting is particularly interesting is that one can see nontrivial links between the QSC and the problem of computing 3-point correlation functions \cite{Cavaglia:2018lxi}. 

%While some components of the construction were understood earlier in \cite{Gromov:2015dfa,Gromov:2016rrp}, most of the results we present were derived only recently in \cite{Gromov:2018lxi}. 

\begin{figure}[ht]
\begin{center}
\includegraphics[scale=0.7]{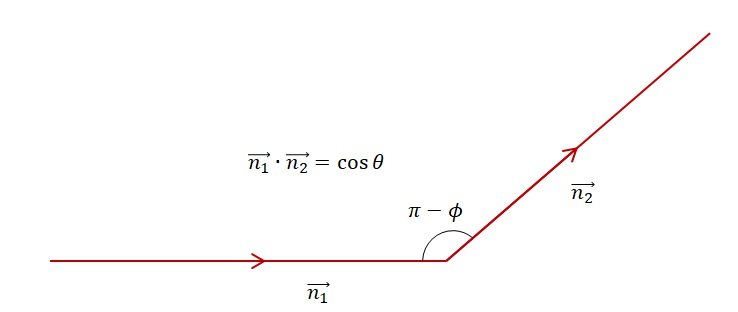}
\end{center}
\caption{\label{fig:cusp} The cusped Wilson line in $\cN=4$ SYM.}
\end{figure}

The cusped Wilson line in $\cN=4$ SYM is an example of an observable outside the spectral problem of local operators which is nevertheless integrable. We will consider a setup with two intersecting Wilson-Maldacena lines, see figure \ref{fig:cusp}. They are defined as
\beq
W={\rm Pexp}  \int\,d\tau  \(i A_\mu \dot x^\mu +\Phi^a n^a  | \dot x |\) 
\eeq
where $n^a$ is a unit six-vector which defines a coupling to the six scalars of $\cN=4$ SYM. On the two lines these vectors are taken to be constant, with a relative angle $\theta$ between them so that  $\vec{n_1}\cdot\vec{n_2}=\cos \theta$. One more parameter in the problem is the geometrical angle $\phi$ between the two lines.

Due to the presence of the cusp, the expectation value of this Wilson line is divergent, with the divergence controlled by a function $\Delta(\lambda,\phi,\theta)$, known as the cusp anomalous dimension\footnote{More precisely, the standard cusp anomalous dimension in our notation is given by $\Gamma_{\rm cusp}=-\Delta$}, which in this case is an analog of the conformal dimension for local operators. In \cite{Correa:2012hh,Drukker:2012de} it was realized that this observable is integrable and $\Delta$ can be computed from a set of TBA equations. Then in \cite{Gromov:2015dfa} the QSC framework was generalized to this observable which made possible its efficient calculation in various regimes.

In the QSC for this setup we still have the same set of Q-functions and QQ relations. One thing that changes is the large $u$ asymptotics. The angles $\phi$ and $\theta$ play the role of boundary twists in the integrability description, and similarly to the spin chain case (see \eq{qexp}) they appear in the exponents at large $u$,
\beq
\label{astwp}
	\bP_1\sim e^{+\theta u}u^{-1/2} \ , \ \bP_2\sim e^{-\theta u}u^{-1/2} \ ,\  \bP_3\sim e^{+\theta u}u^{3/2} \ , \ \bP_4\sim e^{-\theta u}u^{3/2} \ ,
\eeq
\beq
\label{astwq}
	\bQ_1\sim e^{-\phi u}u^{\Delta+1/2} \ , \ \bQ_2\sim e^{+\phi u}u^{\Delta+1/2} \ ,\  \bQ_3\sim e^{-\phi u}u^{-\Delta+1/2} \ , 
	\ \bQ_4\sim e^{+\phi u}u^{-\Delta+1/2} \ ,
\eeq
Here we do not write the constant overall factors in front of the asymptotics, they can be found in appendix~\ref{app:qsc} along with details of the parametrization of the $\bP$-functions in terms of $x(u)$.

Moreover, the Q-functions have an extra branch cut from $u=0$ to $u=\infty$, but once we divide them by $\sqrt{u}$ the cut disappears, so that the functions
\beq
	q_i(u)=\bQ_i/\sqrt{u}
\eeq
have the same analytic structure as before. The last important change is the gluing conditions, which now read
\beqa\la{qtil}
&&\tilde { q}_1(u)={q}_1(-u)\\
&&\tilde { q}_2(u)={ q}_2(-u)\\
&&\tilde {q}_3(u)=a_1\sinh(2\pi u){ q}_2(-u)+{ q}_3(-u)\\
&&\tilde { q}_4(u)=a_2\sinh(2\pi u){ q}_1(-u)+{ q}_4(-u)\;,
\eeqa
where $a_{1,2}$ are some constants.

\subsection{The ladders limit}

We will focus on a simplifying limit in which $\theta\to i\infty$ while the coupling $g=\sqrt{\lambda}/({4\pi})$ goes to zero with the finite combination
\beq
    \hat g=\frac{g}{2} e^{-i\theta/2}\la{lad}
\eeq
playing the role of effective coupling. This limit is known to select only ladder Feynman diagrams and is very similar to the fishnet limit of $\gamma$-deformed $\cN=4$ SYM \cite{Gurdogan:2015csr}.

The QSC simplifies significantly since the coupling goes to zero, which means that the cuts $[-2g,2g]$ disappear and manifest themselves as poles. We can write an expansion for the $\bP$-functions similar to \eq{param}, and since $g\to 0$ one can verify that it becomes a finite Laurent expansion around $u=0$ (we assume that the coefficients $c_{a,n}\sim 1$ in this limit as one can check perturbatively). The 4th order Baxter equation \eq{bax5} can then be written quite explicitly. Nicely, it factorizes and reduces to a 2nd order equation \cite{Cavaglia:2018lxi}
\beq
\label{Bax2p}{
	\left(-2u^2 \cos \phi +2\Delta  u \sin \phi +4 \hat{g}^2\right){q}(u) +u^2
   { q}(u-i)+u^2 { q}(u+i)=0}
\eeq
together with another equation obtained by $\Delta\to-\Delta$. By asymptotics we can identify the solutions of \eq{Bax2p} as
\beq
\label{qaslad}
	q_1\sim e^{u\phi}u^\Delta \ , \ q_4\sim e^{-u\phi}u^{-\Delta}
\eeq
Finally, the gluing condition, which was originally some constraint on the cut $[-2g,2g]$, now becomes a condition on the pole of the Q-function at the origin as one can show \cite{Cavaglia:2018lxi},
\beq
\label{qquant}
{
    	\Delta=-\frac{2\hat g^2}{\sin\phi}
	\frac{q_1(0)\bar q_1'(0)+\bar q_1(0)q_1'(0)}{q_1(0)\bar q_1(0)}  } \;.
\eeq
Thus the problem is reduced to solving the 2nd order equation \eq{Bax2p}, picking the solution $q_1$ with the asymptotics \eq{qaslad} and imposing the gluing condition \eq{qquant} which fixes $\Delta$ in terms of the coupling and $\phi$. 

As an example, at weak coupling $\hat g\to 0$ we can easily solve the Baxter equation and find
\beq
	q_1=e^{u\phi} + \cO(\hat g^2)
\eeq
From \eq{qquant} we then immediately find the scaling dimension $\Delta$ to be
\beq
	\Delta=-4\hat g^2\frac{\phi}{\sin\phi} + \cO(\hat g^4)
\eeq
which reproduces the known 1-loop result. For illustration, at the next order we have \cite{Cavaglia:2018lxi}
\beq
\label{qgs}
q_1 = e^{u\phi } \, \left[ 1 + \hat{g}^2 \, \frac{2 i }{\sin\phi} \,  \left( 2 \, \phi \, (\eta^1_1-  \eta ^{e^{2 i \phi }}_1 )  -  (\eta^1_2-\eta ^{e^{2 i \phi }}_2 )\right) \right] +\cO(\hat g^4) \ 
\eeq
which is written in terms of $\eta$-functions that typically appear in the weak coupling solutions of QSC \cite{Leurent:2013mr,Marboe:2014gma,Gromov:2015dfa} and are defined as
\beq
\label{defeta}
\eta_{s_1,\dots,s_k}^{z_1,\dots,z_k}(u)\equiv \sum_{n_1 > n_2 > \dots > n_k \geq 0}\frac{z_1^{n_1}\dots z_k^{n_k}}{(u+i n_1)^{s_1}\dots (u+i n_k)^{s_k}} \ .
\eeq
We see that $q_1(u)$ has poles at $u=-in$ (with $n=1,2,\dots$) which are remnants of the cuts $[-2g-in,2g-in]$ of the Q-functions at finite coupling.

\subsection{From the QSC to correlators: example}

A longstanding open problem is to find out whether the QSC can provide a new approach to computing 3-point structure constants in $\cN=4$ SYM. The reason to expect a connection is that computation of correlators has close links with wavefunctions of the states, and the wavefunctions in separated variables are expected to factorize into a product of Q-functions which are given by QSC at finite coupling,
\beq
	\Psi \sim Q(x_1)Q(x_2)\dots Q(x_n)
\eeq
The correlator roughly speaking corresponds to a certain overlap of the states, and thus we expect it to be given by a kind of scalar product between the Q-functions. Of course, the details of these connections remain to be uncovered, and are nontrivial even for spin chains. For example, the scalar product in separated variables for higher rank models was found only very recently \cite{Cavaglia:2019pow}.

A natural starting point in this program are the structure constants between two identical operators and the Lagrangian $\<\cO\cO\cL\>$, which are given by the variation of the scaling dimension w.r.t. the coupling $\d\Delta/\d g$ (see \cite{Costa:2010rz}). Thus they establish a link between the spectral problem and more general correlators. 

At present it is not yet known how to compute $\d\Delta/\d g$ in closed form for local operators (in terms of Q-functions evaluated at one given value of the coupling). However, for the cusped Wilson line setup this was done in \cite{Cavaglia:2018lxi} in the ladders limit which we are considering. Even more importantly, it was found there that the structure constant defined by a correlator of three cusps takes a very simple form as essentially an integral of the corresponding three Q-functions. While this calculation is rather involved, here we will describe the much simpler computation of $\d\Delta/\d g$, which already reveals a highly intriguing structure.

Let us first write the difference operator defining the Baxter equation \eq{Bax2p} in the form
\beq
\hat O \equiv \frac{1}{u}\left[
(4\hat g^2-2u^2\cos\phi+2\Delta u\sin\phi)
+u(u-i)D^{-1}+u(u+i)D
\right]\frac{1}{u} 
\eeq
with $D$ being the operator of shift by $i$, so that the Q-functions satisfy
\beq
\hat O q(u)=0\;.
\eeq
It will be enough for us to work with only $q_1$ rather than other Q's, so we will omit the index of the Q-function.

The main observation is that this operator has the following `self-adjointness' property:
\beq
\label{osa}
\vint q^{(A)}(u)\hat O q^{(B)}(u) du=
\vint q^{(B)}(u)\hat O q^{(A)}(u) du\;\;,\;\;\vint\equiv \int_{c-i\infty}^{c+i\infty} \  \  \ .
\eeq
where $c>0$ and $A,B$ label two different functions\footnote{note that $A,B$ are not multi-indices like in section \ref{sec:Qs}} with the same analytic structure as $q_1$. In order to verify \eq{osa}, let us focus on one term first, for which we have
\beq
\vint q^{(A)}(u)u(u+i) Dq^{(B)}(u) du=
\vint q^{(A)}(u)u(u+i) q^{(B)}(u+i) du=
\vint q^{(B)}(u)u(u-i)D^{-1} q^{(A)}(u) du
\eeq
where we shifted the integration variable by $i$ at the second step. We clearly recognize here a part of $\hat O$ acting now on $q^{(A)}$. Treating similarly the remaining terms, we see that \eq{osa} holds. 

The property \eq{osa} now allows us to simply apply the usual perturbation theory logic. Let us consider a small variation $\hat g\to \hat g+\delta \hat g$ of the coupling, then the Q-function will also change and we have
\beq
(\hat O+\delta \hat O)(q+\delta q)=0\;\;,\;\;\delta \hat O = \frac{1}{u^2} (8\hat g\delta \hat g+2u\sin\phi \delta \Delta
)\;.
\eeq
Multiplying this by $q$ and take the integral, we find to first order in the variation
\beq
0=\vint q(\hat O+\delta \hat O)(q+\delta q)du=\vint q\delta\hat O q \;du + \vint q\hat O\delta q\; du =\vint q\delta\hat O q du
\eeq
where at the last step we got rid of the term with $\delta q$ by using the self-adjointness of $\hat O$ to act with it on the left. As a result we get
\beq
	\vint q\delta\hat O q du=0
\eeq
or explicitly
\beq
\vint q (8\hat g\delta \hat g+2u\sin\phi \;\delta \Delta
) q \frac{du}{u^2} = 0 \ ,
\eeq
which means that
\beq
\label{eq:dddg}\frac{\d \Delta}{\d \hat g}=-\frac{4\hat g}{\sin\phi}\frac{\vint\frac{q^2}{u^2} du}
{\vint\frac{q^2}{u} {du}}
\eeq
Finally, introducing the bracket\footnote{the overall prefactor in the bracket is irrelevant and is chosen to be as in \cite{Cavaglia:2018lxi} for consistency}
 for functions with large $u$ asymptotics $f\sim e^{u\beta}u^\alpha$
\beq
\la{eq:thebracket}
\br{f(u) }\equiv \(2\sin\frac{\beta}{2}\)^\alpha\vint f(u)\frac{du}{2\pi i u}\;\;,\;\;c>0\;
\eeq
we obtain
	\beq
\boxed{
\frac{\partial \Delta}{\partial\hat g^2}=-4\frac{\br{q^2\frac{1}u}}{\br{q^2}}
}
\eeq
The interpretation of this result is that in the numerator we have some kind of scalar product where the two Q-functions represent the two $\cO$ operators in the $\<\cO\cO\cL\>$ correlator, while the $1/u$ insertion represents the Lagrangian. In the denominator we have the standard normalization by the 2-pt function,
\beq
	\frac{\partial \Delta}{\partial\hat g^2} \propto C_{123}\propto\frac{\<\cO\cO\cL\>}{\sqrt{\<\cO\cO\>}\sqrt{\<\cO\cO\>}}
\eeq
which for our diagonal correlator gives the norm squared of $q$. Remarkably, a very similar structure was found in \cite{Cavaglia:2018lxi} for the much more complicated correlator of three cusps with scalar insertions (see also \cite{McGovern} for a recent extension to more general insertions).

Let us note that to a given value of the coupling there correspond several values of $\Delta$, associated to scaling dimension of the cusp with certain scalar insertions (see \cite{Cavaglia:2018lxi} for details). A curious consequence of the self-adjointness property is that Q-functions associated to these different states (at fixed coupling) are orthogonal to each other under our bracket as
\beq
0=\vint q^{(A)}(u)(\hat O^{(A)}-\hat O^{(B)})q^{(B)}(u) du =
(\Delta^{(A)}-\Delta^{(B)})2\sin\phi\vint \frac{q^{(A)}(u)q^{(B)}(u)}{u} du \ ,
\eeq
so that for distinct states with $\Delta^{(A)}\neq\Delta^{(B)}$ we have
\beq
	\br{q^{(A)}q^{(B)}}=0
\eeq
This supports the interpretation of this bracket as a natural scalar product for Q-functions.

Recently similar techniques were developed for non-compact spin chains in \cite{Cavaglia:2019pow}, extending the Sklyanin scalar product in SoV beyond simplest $gl(2)$-type models.

\subsection{QSC gluing conditions as finiteness of norm}
\label{sec:glnorm}

Having interpreted the bracket as a scalar product, one may wonder if the states which have a finite norm are singled out in some natural way. Indeed this is the case, and one can show \cite{Cavaglia:2018lxi} that finiteness of the norm $\sqrt{\br{q^2}}$ is in fact equivalent to the rather non-trivial gluing condition \eq{qquant}. In other words, we can simply select physical solutions of the Baxter equation by requiring them to have finite norm, just as in quantum mechanics!

For the cusp in the ladders limit one can in fact resum the diagrams contributing to $\Delta$ by using Bethe-Salpeter methods \cite{Erickson:1999qv,Erickson:2000af,Correa:2012nk}. As a result we get a Schrodinger equation
\beq
4\[-\d_x^2-
\frac{2\hat{g}^2}{ \cosh x + \cos\phi}\]
\psi(x)={E} \psi(x)\;
\eeq
whose bound state energies give the values of $\Delta$ according to $\Delta=-\sqrt{-E}$. It was found in \cite{Cavaglia:2018lxi} that one can construct an explicit map (similar to a Mellin transform) between the Schrodinger and the Baxter equations which establishes their equivalence. This is an example where one can find a direct link between the diagrams and the QSC. Not surprisingly, the norm of the wavefunction translates to the norm of the Q-functions defined by our bracket $\br{q^2}$, which confirms that finiteness of the latter selects the physical solutions.

This suggests in general a new point of view on the gluing conditions, which are the key part of the QSC. They may be interpreted as normalizability with respect to a certain scalar product (associated with Sklyanin's separated variables). We present some speculations related to this in the conclusions section.

\section{Conclusion}

Let us conclude by discussing several of the open problems associated with the QSC and its applications.

It still remains to generalize the QSC to lower dimensional dualities such as AdS$_3$/CFT$_2$ \cite{Sfondrini:2014gza}. The complication there is the presence of massless modes which makes the branch cut structure nontrivial to anticipate. In addition, there are some rather nontrivial deformations of $\cN=4$ SYM for which integrability persists but QSC is not yet known (e.g. the dipole deformation \cite{Guica:2017jmq} and more general Yang-Baxter deformations).

Derivation of exact spectral equations such as QSC for AdS/CFT has been mainly guided by the string picture, where one invokes standard 2d techniques such as bootstrapping the exact S-matrix and so on, leading finally to TBA/Y-system and then QSC. In this context, the fishnet model certainly sheds new light on the QSC, especially with the very recent derivation of its dual string-like model starting from Feynman graphs \cite{Gromov:2019aku,Gromov:2019bsj,Gromov:2019jfh}. That dual theory can be quantized essentially from first principles and one can truly derive the Baxter equations which were known before from QSC. That exciting development leads one to speculate that even for $\cN=4$ SYM the integrability could one day be proven rigorously and in particular the QSC could be derived from gauge theory.

Yet another direction which is under active investigation is the application of QSC to correlators. It is being explored together with further developing the separation of variables program in general, especially for higher-rank models where little was known until recently \cite{Sklyanin:1992sm,Smirnov2001,Gromov:2016itr,Cavaglia:2019pow,DimaToapp} (see also \cite{Maillet:2018czd,Maillet:2019ayx,SmirnovQuantM,Gromov:2018cvh,Derkachov:2018ewi}). It has been found in several different settings that the QSC can simplify the results for 3-pt structure constants, drastically in some cases. One may expect that a similar picture should hold in the full $\cN=4$ SYM theory, and we are already starting to see it explicitly for the simpler fishnet model where compact results for $\<\cO\cO\cL\>$ correlators can be derived in terms of Q-functions \cite{preparation} (where we first add an extra spacetime twist to the fishnet model, see \cite{TwistingFishing}, and it can be removed at the end). The reformulation of gluing conditions as a finiteness of norm discussed in sec \ref{sec:glnorm} is also likely to uplift to full $\cN=4$ SYM as suggested by derivation of the gluing conditions in \cite{Gromov:2015vua}. Moreover, very recently some correlators of determinant operators were expressed in TBA language \cite{Jiang:2019zig,Jiang:2019xdz},  suggesting again they may be fruitfully reformulated using the QSC. One may hope that if this program is successful, it would allow one to obtain correlation functions with the same precision and efficiency as the spectrum, and thus complete the solution of a 4d gauge theory for the first time.

\section*{Acknowledgements}
I thank A.~Cavaglia, N.~Gromov, G.~Sizov and S.~Valatka for collaboration on related subjects. I am also grateful to C.~Marboe and D.~Volin for sharing some lecture notes, to V.~Kazakov for discussions, and to A.~Liashyk for comments on the text. My work is supported by Agence Nationale de la Recherche LabEx grant ENS-ICFP ANR-10-
LABX-0010/ANR-10-IDEX-0001-02 PSL.

	\appendix
	
\section{Details on the QSC}
\label{app:qsc}

The coefficients entering the fourth order Baxter type equation \eq{bax5} on $\bQ_i$ are given by:
\beqa
D_0&=&{\rm det}
\(
\bea{llll}
\bP^{1[+2]}&\bP^{2[+2]}&\bP^{3[+2]}&\bP^{4[+2]}\\
\bP^{1}&\bP^{2}&\bP^{3}&\bP^{4}\\
\bP^{1[-2]}&\bP^{2[-2]}&\bP^{3[-2]}&\bP^{4[-2]}\\
\bP^{1[-4]}&\bP^{2[-4]}&\bP^{3[-4]}&\bP^{4[-4]}
\eea
\)\;,\\
D_1&=&{\rm det}
\(
\bea{llll}
\bP^{1[+4]}&\bP^{2[+4]}&\bP^{3[+4]}&\bP^{4[+4]}\\
\bP^{1}&\bP^{2}&\bP^{3}&\bP^{4}\\
\bP^{1[-2]}&\bP^{2[-2]}&\bP^{3[-2]}&\bP^{4[-2]}\\
\bP^{1[-4]}&\bP^{2[-4]}&\bP^{3[-4]}&\bP^{4[-4]}
\eea
\)\;,\\
D_2&=&{\rm det}
\(
\bea{llll}
\bP^{1[+4]}&\bP^{2[+4]}&\bP^{3[+4]}&\bP^{4[+4]}\\
\bP^{1[+2]}&\bP^{2[+2]}&\bP^{3[+2]}&\bP^{4[+2]}\\
\bP^{1[-2]}&\bP^{2[-2]}&\bP^{3[-2]}&\bP^{4[-2]}\\
\bP^{1[-4]}&\bP^{2[-4]}&\bP^{3[-4]}&\bP^{4[-4]}
\eea
\)\;,\\
\bar D_1&=&{\rm det}
\(
\bea{llll}
\bP^{1[-4]}&\bP^{2[-4]}&\bP^{3[-4]}&\bP^{4[-4]}\\
\bP^{1}&\bP^{2}&\bP^{3}&\bP^{4}\\
\bP^{1[+2]}&\bP^{2[+2]}&\bP^{3[+2]}&\bP^{4[+2]}\\
\bP^{1[+4]}&\bP^{2[+4]}&\bP^{3[+4]}&\bP^{4[+4]}
\eea
\)\;,\\
\bar D_0&=&{\rm det}
\(
\bea{llll}
\bP^{1[-2]}&\bP^{2[-2]}&\bP^{3[-2]}&\bP^{4[-2]}\\
\bP^{1}&\bP^{2}&\bP^{3}&\bP^{4}\\
\bP^{1[+2]}&\bP^{2[+2]}&\bP^{3[+2]}&\bP^{4[+2]}\\
\bP^{1[+4]}&\bP^{2[+4]}&\bP^{3[+4]}&\bP^{4[+4]}
\eea
\)\;.
\eeqa

For the twisted case discussed in section~\ref{sec:correl} (see \cite{Gromov:2015dfa} for all details), the $\bP$-functions can be parametrized as
\beqa
\label{cuspas}
\bP_1(u)&=&+\epsilon\;u^{1/2}\; e^{+\theta u}\; {\bf f}(+u)\ ,\\
\nn
\bP_2(u)&=&-\epsilon\;u^{1/2}\; e^{-\theta u}\; {\bf f}(-u)\ ,\\
\nn
\bP_3(u)&=&+\epsilon\;u^{1/2}\; e^{+\theta u}\; {\bf g}(+u)\ ,\\
\nn
\bP_4(u)&=&+\epsilon\;u^{1/2}\; e^{-\theta u}\; {\bf g}(-u)\ .
\eeqa
where the functions ${\bf f}(u)$ and ${\bf g}(u)$ have powerlike asymptotics at large $u$ with ${\bf f}\simeq 1/u$ and ${\bf g}\simeq u$. The prefactor $\epsilon$ in this normalization reads
\beq
    \epsilon=\sqrt{\frac i2}\;
\frac{\cos \theta -\cos \phi}{\sin \theta}
\;.
\eeq

In terms of $x$ defined in \eq{defx} these functions are power series,
\beqa
\label{fgAB}
{\bf f}(u)=
\frac{1}{gx}+\sum_{n=1}^\infty \frac{g^{n-1}A_n}{x^{n+1}}
\;\;,\;\;
{\bf g}(u)=
\frac{u^2+B_0u}{gx}+\sum_{n=1}^\infty  \frac{g^{n-1}B_n}{x^{n+1}}\;.
\eeqa
The coefficients $A_n$ and $B_n$ contain information about the AdS conserved charges including $\Delta$. In particular,
\beqa
\label{epsab0}
A_1g^2-B_0&=&-\frac{ 2 \cos \theta  \cos \phi+\cos (2 \theta) -3}{2 \sin\theta (\cos \theta -\cos \phi)}\;,\\
\nn
	\Delta^2&=&
	\frac{(\cos\theta-\cos\phi)^3}{\sin\theta\sin^2\phi}\[
A_3 g^6+\frac{A_1^2 g^4 (1-\cos\theta \cos\phi)}{
\sin\theta(\cos \theta -\cos\phi)}-A_2 g^4 \cot\theta
\right.\\
\label{deltaas0}&&   \left.
-g^2 \left(B_0+B_1+\cot \theta\right)
-A_1 g^2
   \left(A_2 g^4-2 g^2+\frac{1}{\sin^2\theta}\right)
\]\;.
\eeqa

\end{document}